\documentclass{article}
\newcommand{\proglang}[1]{\textbf{#1}}
\newcommand{\pkg}[1]{\textbf{#1}}

\usepackage{graphicx, float, amsmath,mathrsfs, hyperref,color,listings}
\usepackage{amssymb}
\usepackage{upquote} %allows the ' symbol to be copy pasted from verbatim into code windows
\usepackage{setspace}
\usepackage{moreverb}
\usepackage{natbib}
\usepackage[mathletters]{ucs}
\usepackage[utf8x]{inputenc}
\usepackage[a4paper]{geometry}

\newcommand{\htx}[2]{\hspace{ #1 cm} \text{ #2 } }

\title{\pkg{interAdapt} – An Interactive Tool for Designing and Evaluating Randomized Trials with Adaptive Enrollment Criteria}
\date{}
\author{Aaron Fisher, Harris Jaffee, \& Michael Rosenblum}
\begin{document}
\maketitle

\begin{abstract}
The \pkg{interAdapt} \proglang{R} package is designed to be used by
statisticians and clinical investigators to plan randomized trials. It
can be used to determine if certain adaptive designs offer tangible
benefits compared to standard designs, in the context of investigators'
specific trial goals and constraints. Specifically, \pkg{interAdapt}
compares the performance of trial designs with adaptive enrollment
criteria versus standard (non-adaptive) group sequential trial
designs. Performance is compared in terms of power, expected trial
duration, and expected sample size. Users can either work directly in
the \proglang{R} console, or with a user-friendly \pkg{shiny} application that requires
no programming experience. Several added features are available when
using the \pkg{shiny} application. For example, the application allows
users to immediately download the results of the performance
comparison as a csv-table, or as a printable, html-based report.
\end{abstract}

\section*{Introduction} \label{sec:intro}
Group sequential, randomized trial designs involve rules for early stopping of an entire trial based on analyses of accrued data. Such early stopping could occur if there is strong evidence early in the trial of benefits or harms of the new treatment being studied. Adaptive enrichment designs involve rules for restricting enrollment criteria based on data accrued in an ongoing trial. For example, enrollment could be stopped for a certain subpopulation if there is strong early evidence that the treatment does not benefit that group.
We focus on the  class of designs introduced by \cite{Rosenblum2013AdaptMISTIE}, which combines features of both group sequential and adaptive enrichment designs. For conciseness, we refer to designs in this class as  ``adaptive designs.''
These are contrasted with ``standard designs,'' 
defined to  be group sequential designs where the enrollment criteria cannot be changed during the trial (except that the entire trial may be stopped early for efficacy or futility). 

We introduce the \pkg{interAdapt} \proglang{R} package, a user friendly set of tools for exploring certain types of adaptive enrichment designs. The package contains a densely featured \pkg{shiny} application, ideal for users with little to no \proglang{R} programming experience, as well as an \proglang{R} function (\verb!compute_design_performance!) which provides the same core computational functionality that underlies the \pkg{shiny} application. Several input parameters are available to allow the user to describe the context of his/her trial. Computations for generating output typically require less than 15 seconds on a standard commercial laptop. 

Several added features are available when using the \pkg{shiny} application, as opposed to the command line function. For example, users can upload data from previous studies, and the application will automatically compute the relevant parameters for the trial being planned. Once entered, the full set of input parameters can be saved to the user's computer for use in future sessions. Results of the design comparisons can be immediately downloaded in the form of either csv-tables, or printable, html-based reports. The \pkg{shiny} application is also hosted on the RStudio webserver, and can be accessed online without installing the \proglang{R} programming language.

To demonstrate our software, we consider the problem of 
 planning a Phase III trial for a new surgical treatment of stroke, which is considered by \cite{Rosenblum2013AdaptMISTIE}.
The new treatment is  called Minimally-Invasive Surgery Plus rt-PA for Intracerebral Hemorrhage (MISTIE), and is described in detail by \cite{MISTIE_prelim2008}. 
Previous trials had almost exclusively enrolled participants with little or no intraventricular hemorrhage (IVH) at baseline (referred to as small IVH participants).
However, it was conjectured that the treatment may also benefit  participants with large IVH volume at baseline.  
The goal of the Phase III trial being planned was to determine whether MISTIE is effective for the combined population of those with small or large IVH, and, if not, to determine whether MISTIE is effective for the small IVH population (for whom there was greater prior evidence). A  standard trial design  may be inefficient at simultaneously answering these questions.
An alternative is to use an adaptive trial design that  first recruits from the combined population, and then decides whether to restrict enrollment based on results from interim  analyses. 
Though we focus on this stroke trial application throughout, our software tool can be applied in many disease areas.

In Section \ref{sec:problemDescription}, we  formally define the hypothesis testing problem to be addressed by the different trial designs. In Section \ref{ADDPLAN}, we compare our software to the most similar, currently available software, AptivSolutions ADDPLAN PE (Participant Enrichment), and the \pkg{asd} \proglang{R} package. 
In Section \ref{sec:running-interAdapt}, we describe how to install \pkg{interAdapt} on a personal computer, run the \pkg{shiny} application locally, use the package functionality in the command line, and access the application online through a web browser. Section \ref{sec:UI} describes the inputs available when using \pkg{interAdapt}, and discusses the interpretation of the package's output. In Section \ref{sec:example}, we present an example demonstrating how an adaptive design is created and analyzed with \pkg{interAdapt}. 

\section{Problem description}
\label{sec:problemDescription}

We consider the problem of designing a randomized trial to test whether a new treatment is superior to control, for a given population (e.g., those with intracerebral hemorrhage in  the MISTIE example).
Consider the case where we have two subpopulations, referred to as subpopulation $1$ and subpopulation $2$, which partition the overall population of interest. These must be specified before the trial starts, and be defined in terms of participant attributes measured at baseline (e.g., having a high initial severity of disease or a certain biomarker value). 
We focus on situations where  there is suggestive, prior evidence that the treatment may be more likely to benefit subpopulation $1$.
In the MISTIE trial example, subpopulation 1 refers to small IVH participants, and subpopulation 2 refers to large IVH participants. 
Let $π_1$ and $π_2$ denote the proportion of the population in subpopulations 1 and 2, respectively.

Both the adaptive and standard designs discussed here involve enrollment over time, and include predetermined rules for stopping the trial early based on interim analyses. Each trial consists of $K$ stages, indexed by $k$. 
In stages where both subpopulations are enrolled, we assume that the proportion of newly recruited participants  in each subpopulation $s \in \{1,2\}$ is equal to the corresponding population proportion $\pi_s$.

For a given design, let $n_k$ denote the maximum number of participants to be enrolled during stage $k$. The number enrolled during stage $k$ will be less than $n_k$ if the trial is entirely stopped before stage $k$ (so that no participants are enrolled in stage $k$) or if in the adaptive design enrollment is restricted to only subpopulation 1 before stage $k$ (as described in Section~\ref{sub:decisionRules}). For each subpopulation $s \in \{1,2\}$ and stage $k$, let $N_{s,k}$ denote the maximum cumulative number of subpopulation $s$ participants who have enrolled by the end of stage $k$. Let $N_{C,k}$ denote the maximum cumulative number of enrolled participants from the combined population by the end of stage $k$, i.e.,  $N_{C,k}=N_{1,k}+N_{2,k}$.
The sample sizes will generally differ for different designs.

Let $Y_{i,k}$ be a binary outcome variable for the $i^{th}$ participant recruited in stage $k$, where $Y_{i,k}=1$ indicates a successful outcome. Let $T_{i,k}$ be an indicator of   the $i^{th}$ participant recruited in stage $k$ being assigned to the treatment. We assume for each participant that there is an equal probability of being assigned to  treatment ($T_{i,k}=1$) or control $(T_{i,k}=0$), independent of the participant's subpopulation. We also assume outcomes are observed very soon after enrollment, so that all outcome data is available from currently enrolled participants at each interim analysis.

For subpopulation $1$, denote the probability of a successful outcome under treatment as $p_{1t}$, and the probability of a successful outcome under control as $p_{1c}$. Similarly, for subpopulation $2$, let $p_{2t}$ denote the probability of a success under treatment, and $p_{2c}$ denote the probability of a success under control. 
We assume each of $p_{1c},p_{1t},p_{2c},p_{2t}$ is in the interval $(0,1)$.
We define the true average treatment effect for a given population to be the difference in the probability of a successful outcome comparing treatment versus control.

In the remainder of this section we give an overview of the relevant concepts needed to understand and use \pkg{interAdapt}. A more detailed discussion of the theoretical context, and of the efficacy boundary calculation procedure, is provided by \cite{Rosenblum2013AdaptMISTIE}.

\subsection{Hypotheses}
\label{sub:hypotheses}

We focus on testing the null hypothesis that, on average, the treatment is no better than control for subpopulation $1$, and the analogous null hypothesis for the combined population. Simultaneous testing of null hypotheses for these two populations was also the goal for the two-stage, adaptive enrichment designs of \cite{wangetal2007}.
We define our two null hypotheses, respectively, as

\begin{itemize}
\item $H_{01}$: $p_{1t}-p_{1c}≤0$;%The treatment effect in subpopulation $1$ is less than or equal to zero.
\item $H_{0C}$: $π_1(p_{1t}-p_{1c}) + π_2(p_{2t}-p_{2c}) ≤ 0$. %The average treatment effect in the combined population is less than or equal to zero.	
\end{itemize}

\pkg{interAdapt} compares different designs for testing these null hypotheses. 
An adaptive design testing both null hypotheses (denoted $AD$) is compared to two standard designs. The first standard design, denoted $SC$, enrolls the combined population and only tests $H_{0C}$. The second standard design, denoted $SS$, only enrolls subpopulation 1 and tests $H_{01}$.
All three trial designs consist of $K$ stages; the decision to entirely stop the trial early can be made at the end of any stage, based on a preplanned rule. The trials differ in that $SC$ and $SS$ never change their enrollment criteria, while $AD$ may switch from enrolling the combined population  to enrolling only participants from subpopulation $1$.

The standard designs discussed here are not identical to those discussed in Section 6.1 of \citep{Rosenblum2013AdaptMISTIE}, which test both hypotheses simultaneously. Implementing standard designs such as those discussed in \citep{Rosenblum2013AdaptMISTIE} into the \pkg{interAdapt} software is an area of future research.

Though it is not of primary interest, we occasionally refer below to the global null hypothesis, defined  to be that $p_{1t}-p_{1c}=p_{2t}-p_{2c}=0$, i.e., zero mean treatment effect in both subpopulations.

%Old Note: Whenever any of the trials $AD$, $SC$ or $SS$ is stopped early, there will be some participants who have been enrolled but who’s outcomes have not yet been measured. These participants are referred to as ''overrunning'' or ''pipeline'' participants. \pkg{interAdapt} currently discards measurements from these overruning participants in the final analysis. Incorporating these measurements is a goal for future work.
%AF note to self: There are a few ways to do this, such as bayesian probability of passing a threshold once the participants come in. There are also ways to set "double thresholds," one threshold for when to stop, and one (usually lower) threshold for whether or not to declare significant results once the overrunning participants are measured.

\subsection{Test statistics}
\label{sub:testStats}
Three (cumulative) z-statistics are computed at the end of each stage $k$. The first is based on all enrolled participants in the combined population, the second is based on all enrolled participants in subpopulation 1, and the third is based on all enrolled participants in subpopulation 2.  Each z-statistic is a standardized difference in sample means, comparing outcomes in the treatment arm versus the control arm.
Let $Z_{C,k}$ denote the z-statistic for the combined population at the end of stage $k$, which  takes the following form:

\[\begin{split}
Z_{C,k}&=\left[
\frac{\sum_{k'=1}^k \sum_{i=1}^{n_{k'}}Y_{i,k'}T_{i,k'} }
{\sum_{k'=1}^k \sum_{i=1}^{n_{k'}}T_{i,k'}}
-
\frac{\sum_{k'=1}^k \sum_{i=1}^{n_{k'}} Y_{i,k'}(1-T_{i,k'})} 
{\sum_{k'=1}^k \sum_{i=1}^{n_{k'}}(1-T_{i,k'})}
\right] \\
& \htx{1}{} \times
\left\lbrace
\left(     \frac{2}{  N_{C,k}  }       \right)
\left(
\sum_{s ∈ \{ 1,2\}} π_s[p_{sc}(1-p_{sc}) + p_{st}(1-p_{st})]
\right)
\right\rbrace ^{-1/2}
\end{split}\]

The term in square brackets is the difference in sample means between the treatment and control groups. The term in curly braces is the variance of this difference in sample means. $Z_{C,k}$ is only computed at stage $k$ if the combined population has been enrolled up through the end of stage $k$ (otherwise it is undefined). Our designs never use $Z_{C,k}$ after stages where the combined population has stopped being enrolled.
Let $Z_{1,k}$ and $Z_{2,k}$ denote analogous z-statistics restricted to participants in subpopulation $1$ and subpopulation $2$, respectively. These are formally defined in
\citep{Rosenblum2013AdaptMISTIE}.

%Previous version of decision rules.
\iffalse 
The z-statistic for subpopulation 1 can be written as follows, where $A_{i,k}$ is the indicator that the $i ^{th}$ subject recruited in stage $k$ is in subpopulation $1$:
\[\begin{split}
Z_{1,k}&=\left[
\frac{\sum_{k'=1}^k \sum_{i=1}^{n_{k'}}Y_{i,k'}T_{i,k'}A_{i,k'} }
{\sum_{k'=1}^k \sum_{i=1}^{n_{k'}}T_{i,k'}A_{i,k'}}
-
\frac{\sum_{k'=1}^k \sum_{i=1}^{n_{k'}} Y_{i,k'}(1-T_{i,k'})A_{i,k'}} 
{\sum_{k'=1}^k \sum_{i=1}^{n_{k'}}(1-T_{i,k'})A_{i,k'}}
\right] \\
& \htx{1}{} \times
\left\lbrace
\left(     \frac{2}{\sum_{k'=1}^k \sum_{i=1}^{n_{k'}}A_{i,k'}}       \right)
\left(
π_1[p_{1c}(1-p_{1c}) + p_{1t}(1-p_{1t})]
\right)
\right\rbrace ^{-1/2}
\end{split}\]
The z-statistic $Z_{2,k}$ is similar to the above, with each occurrence of $A_{i,k'}$ replaced by $(1-A_{i,k'})$, $p_{1c}$ replaced by $p_{2c}$, and $p_{1t}$ replaced by $p_{2t}$.
The decision rules defined below involve stopping boundaries based on the z-statistics observed at the end of each stage. 
\fi

\subsection{Type I error control}
\label{sub:typeIerror}

The familywise (also called study-wide) Type I error rate is the probability of rejecting one or more true null hypotheses.
For a given design, we say that the familywise Type I error rate is strongly controlled at level $α$ if 
for any values of  $p_{1c},p_{1t},p_{2c},p_{2t}$ (assuming each is in the interval $(0,1)$), 
the probability of rejecting at least one true null hypothesis (among $H_{0C}, H_{01}$) is at most $α$. To be precise, we mean such strong control holds asymptotically, as sample sizes in all stages go to infinity, as formally defined by \cite{Rosenblum2013AdaptMISTIE}. %The reason for this caveat is that we use the asymptotic approximation, common in the design and analysis of randomized trials with binary outcomes, that z-statistics converge in distribution to a normal by the central limit theorem.
For all three designs, $AD$, $SC$, and $SS$, we require the familywise Type I error rate to be strongly controlled at level $α$. 
Since the two standard designs $SS$ and $SC$ each only test a single null hypothesis, the familywise Type I error rate for each design is equal to the  Type I error rate for the corresponding, single hypothesis test.

\subsection{Decision rules for early stopping and for modifying enrollment criteria}
\label{sub:decisionRules}

The decision rules for the standard design $SC$ consist of efficacy and futility boundaries for $H_{0C}$, based on the statistics $Z_{C,k}$. At the end of each stage $k$,  the test statistic $Z_{C,k}$ is calculated. If $Z_{C,k}$ is above the efficacy boundary for stage $k$, the design $SC$ rejects $H_{0C}$ and stops the trial. If $Z_{C,k}$ is between the efficacy and futility boundaries for stage $k$, the trial is continued through the next stage (unless the last stage $k=K$ has been completed). If $Z_{C,k}$ is below the futility boundary for stage $k$, the design $SC$ stops the trial and fails to reject $H_{0C}$. \pkg{interAdapt} makes the simplification that the number of participants $n_k$ enrolled in each stage of $SC$ is a constant, denoted  $n_{SC}$, that the user can set.

The efficacy boundaries for $SC$ are set to be proportional to those described by Wang and Tsiatis (1987). Specifically, the efficacy boundary for the $k^{th}$ stage is set to $e_{SC}(N_{C,k}/N_{C,K})^{\delta}$, where $K$ is the total number of stages, $δ$ is a constant in the range $[-.5,.5]$, and $e_{SC}$ is the constant computed by  \pkg{interAdapt}  to ensure the familywise Type I error rate is at most $\alpha$. Since $n_{k}$ is set equal to $n_{SC}$ for all values of $k$, the maximum cumulative sample size $N_{C,k}$ reduces to $\sum_{k'=1}^k n_{SC}=k n_{SC}$, and the boundary at stage $k$ reduces to the simpler form $e_{SC}(k/K)^\delta$. By default, \pkg{interAdapt} sets $\delta$ to be $-0.5$, which corresponds to the efficacy boundaries of \cite{obrienfleming}. 

In order to calculate $e_{SC}$, \pkg{interAdapt} makes use of the fact that the random vector of test statistics ($Z_{C,1},Z_{C,2},…Z_{C,K}$) converges asymptotically to a multivariate normal distribution with a known covariance structure \citep{JennisonTurnbullBook}. %(Jennison and Turnbull, 1999, Chapter 3)
Using the \pkg{mvtnorm} package \citep{mvtnorm} in \proglang{R} to evaluate the multivariate normal distribution function, \pkg{interAdapt} computes the proportionality constant $e_{SC}$ to ensure the probability of $Z_{C,k}$ exceeding $e_{SC}(N_{C,k}/N_{C,K})^{\delta}$ at one or more stages $k$ is less than or equal to $α$ at the global null hypothesis defined in Section~\ref{sub:hypotheses}.

In $SC$, as well as in $SS$ and $AD$, \pkg{interAdapt} uses non-binding futility boundaries. That is, the familywise Type I error rate is controlled at level α regardless of whether the futility boundaries are adhered to or ignored. The motivation  is that regulatory agencies may prefer non-binding futility boundaries to ensure Type I error control even if a decision is made to continue the trial despite a futility boundary being crossed.

In calculations of power, expected sample size, and expected trial duration, \pkg{interAdapt} assumes futility boundaries are adhered to. 

Futility boundaries for the first $K-1$ stages of $SC$ are set equal to $f_{SC}(N_{C,k}/N_{C,K})^{\delta}$, where $f_{SC}$ is a proportionality constant set by the user. By default, the constant $f_{SC}$ is set to be negative (so the trial  is only stopped for futility  if the z-statistic is below the corresponding negative threshold), although this is not required. In the $K ^{th}$ stage of the trial, \pkg{interAdapt} sets the futility boundary to be equal to the efficacy boundary. This ensures that the final z-statistic $Z_{C,K}$ crosses either the efficacy boundary or the futility boundary.

The decision boundaries for the design $SS$  are defined analogously as for the design $SC$, except using z-statistics $Z_{1,k}$. \pkg{interAdapt} makes the simplification that the number of participants $n_k$ enrolled in each stage $k$ of $SS$ is constant, denoted by $n_{SS}$, and set by the user.
The efficacy boundary for the $k^{th}$ stage is set equal to $e_{SS}(N_{1,k}/N_{1,K})^{\delta}$, where $e_{SS}$ is the constant computed by  \pkg{interAdapt}  to ensure  the  Type I error rate is at most $\alpha$. The first $K-1$ futility boundaries for $H_{01}$ are set equal to $f_{SS}(N_{1,k}/N_{1,K})^{\delta}$,  where $f_{SS}$ is a constant that can be set by the user. The futility boundary in stage $K$ is set equal to the final efficacy boundary in stage $K$.

Consider the adaptive design $AD$.
\pkg{interAdapt} allows the user to a priori specify a final stage  at which there will be a test of  the null hypothesis for the combined population, denoted by stage $k^*$. Regardless of the results at stage $k^*$, $AD$ always stops enrolling from subpopulation $2$ at the end stage $k^*$. This reduces the maximum sample size of $AD$ compared to allowing enrollment from both subpopulations through the end of the trial.
The futility boundaries $l_{2,k}$ are not defined for $k>k^*$, since subpopulation 2 is not enrolled after stage $k^*$. 
The user may effectively turn off the option described in this paragraph by setting $k^*=K$, the total number of stages; then the combined population may be enrolled throughout the trial.

For the $AD$ design, the user can specify the following two types of per-stage sample sizes: one for stages where both subpopulations are enrolled $(k \leq k^*)$, and one for stages where only participants in subpopulation 1 are enrolled $(k > k^*)$. We refer to these two sample sizes as $n^{(1)}$ and $n^{(2)}$, respectively.

%Decision boundaries for $AD$ vary from those of the standard designs two ways. First, 
Because $AD$ simultaneously tests $H_{0C}$ and $H_{01}$ it has two sets of decision boundaries. For the $k^{th}$ stage of $AD$, let $u_{C,k}$ and $u_{1,k}$ denote the efficacy boundaries for $H_{0C}$ and $H_{01}$, respectively. The boundaries $u_{C,k}$ 
 are set equal to $e_{AD,C}(N_{C,k}/N_{C,K})^{\delta}$ for each $k\leq k^*$; 
the boundaries $u_{1,k}$ are set equal to  $e_{AD,1}(N_{1,k}/N_{1,K})^{\delta}$ for each $k \leq K$. 
The constants $e_{AD,C}$  and $e_{AD,1}$ are set such that the probability of rejecting one or more null hypotheses under the global null hypothesis is $\alpha$ (ignoring futility boundaries). It is proved by \cite{Rosenblum2013AdaptMISTIE} that this strongly controls the familywise Type I error rate at level $\alpha$. The algorithm for computing the proportionality constants $e_{AD,C}, e_{AD,1}$ is described later in this section.

%Note to AF: There are several pairs of $e_{AD,C}$  and $e_{AD,1}$ that satisfy this condition for the Type I error rate. For each possible $ e_{AD,C}$ there is a one to one correspondence with a $e_{AD,1}$, which ensures that the probability of rejecting either hypothesis under the global null is zero. 

The boundaries for futility stopping of enrollment from certain population in the $AD$ design, at the end of stage $k$, are denoted by $l_{1,k}$ and $l_{2,k}$. These stopping boundaries are defined relative to the test statistics $Z_{1,k}$ and $Z_{2,k}$, respectively. The boundaries $l_{1,k}$ and $l_{2,k}$ are set equal to $f_{AD,1}(N_{1,k}/N_{1,K})^{\delta}$ (for $k\leq K$) and $f_{AD,2}(N_{2,k}/N_{2,K})^{\delta}$ (for $k < k^*$), respectively, where $f_{AD,1}$ and $f_{AD,2}$ can be set by the user.  In stage $k^*$, the futility boundary $l_{2,k^*}$ is set to ``Inf'' (indicating $\infty$), to reflect that we stop enrollment in subpopulation 2. At the end of each stage, $AD$ may decide to continue enrolling from the combined population, enroll only from subpopulation 1 for the remainder of the trial, or stop the trial entirely.  Specific decision rules based on these boundaries for the z-statistics are described below. 

As described in \citep{Rosenblum2013AdaptMISTIE}, the decision rule in $AD$ consists of the following steps carried out at the end of each stage $k$:

\begin{description}
\item 1. (Assess Efficacy) 
 If $Z_{1,k}>u_{1,k}$, reject $H_{01}$.
   If $k\leq k^*$ and  $Z_{C,k} > u_{C,k}$, reject $H_{0C}$. 
 If $H_{01}$, $H_{0C}$, or both  are rejected, stop all enrollment and end the trial.
\item 2. (Assess Futility of Entire Trial) Else, if $Z_{1,k} ≤ l_{1,k}$ or if this is the final stage of the trial, stop all enrollment and end the trial for futility, failing to reject  any null hypothesis.
\item 3. (Assess Futility for $H_{0C}$) Else, if $Z_{2,k} ≤ l_{2,k}$, or if $k\geq k^*$, stop enrollment from subpopulation $2$ in all future stages. In this case, the following steps are iterated at each future stage:
	\subitem  3a. If $Z_{1,k} > u_{1,k}$, reject $H_{01}$ and stop all enrollment.
	\subitem  3b. If $Z_{1,k} ≤ l_{1,k}$ or if this is the final stage of the trial, fail to reject any null hypothesis  and stop all enrollment.
	\subitem  3c. Else, continue enrolling from only subpopulation $1$. If $k < k^*$ then $π_1n^{(1)}$ participants from subpopulation 1 should be enrolled in the next stage. If $k \geq k^*$, then $n^{(2)}$ participants from subpopulation 1 should be enrolled in the next stage. In all future stages, ignore steps 1, 2, 4, and use steps 3a--3c.
\item  4. (Continue Enrollment from Combined Population) Else, continue by enrolling $\pi_1 n^{(1)}$ participants from subpopulation 1 and $\pi_2 n^{(1)}$ participants from subpopulation 2 for the next stage.
\end{description}

The motivation for Step 2 is that there is assumed to be prior evidence that if the treatment works, it will work for subpopulation 1. Therefore, if subpopulation 1 is stopped for futility, the whole trial is stopped. It is an area of future research to consider modifications to this rule, and to incorporate testing of a null hypothesis for only subpopulation 2.

A consequence of the rule in Step 3 is that Steps 1, 2, and 4 are only carried out for stages $k\leq k^*$.  This occurs since 
 Step 3 restricts enrollment to subpopulation 1 if $Z_{2,k} ≤ l_{2,k}$ or  $k\geq k^*$, and if so runs Steps 3a--3c through the remainder of the trial.

We next describe the algorithm used by  \pkg{interAdapt} to compute the proportionality constants $e_{AD,C}, e_{AD,1}$ that define the efficacy boundaries $u_{C,k},u_{1,k}$. These are selected to ensure the familywise Type I error rate is strongly controlled at level $\alpha$. By Theorem~5.1 of  \citep{Rosenblum2013AdaptMISTIE}, to guarantee such strong control of the familywise Type I error rate, it suffices to set $u_{C,k},u_{1,k}$ such that the familywise Type I error rate is at most $\alpha$ at the global null hypothesis defined in Section~\ref{sub:hypotheses}.
The algorithm takes as input the following, which are set by the user as described in Section~\ref{sub:basic-params}: the per-stage sample sizes $n^{(1)},n^{(2)}$, the study-wide (i.e., familywise) Type I error rate $\alpha$, and a value $a_c$ in the interval $[0,1]$. 
Roughly speaking, $a_c$ represents the fraction of the study-wide Type I error $\alpha$ initially allocated to testing $H_{0C}$, as described next.

The algorithm temporarily sets $e_{AD,1}= \infty$ (effectively ruling out rejection of $H_{01}$)
and computes (via binary search) the smallest value $e_{AD,C}$ such the probability of rejecting $H_{0C}$ is $a_c α$ under the global null hypothesis defined in Section~\ref{sub:hypotheses}. This defines $e_{AD,C}$. 
Next,  \pkg{interAdapt} computes the smallest constant $e_{AD,1}$ such that the probability of rejecting at least one null hypothesis under the global null hypothesis  is at most $\alpha$. 

\iffalse
\begin{equation}
\Prob \left(
Z_{C,k}>e_{AD,C} \left\{\frac{\sum_{k'=1}^{K} n_{k'}}{\sum_{k'=1}^{k}n_{k'}}\right\}^{-δ} \text{  or  } 
Z_{1,k}> e_{AD,1}\left\{\frac{\sum_{k'=1}^{K} n_{k'}}{\sum_{k'=1}^{k}n_{k'}}\right\}^{-δ}\text{  for any $k$}.
\right) ≤ α  \nonumber.
\end{equation}
\fi
All of the above computations use the approximation, based on the multivariate central limit theorem, that the joint distribution of the  z-statistics is multivariate normal  with covariance matrix as given, e.g., by \cite{JennisonTurnbullBook,Rosenblum2013AdaptMISTIE}.

%OlD RULES
% \begin{description}
% \item (1) Assess Efficacy in the Combined Population: If $Z_{C,k} > u_{C,k}$, reject $H_{0C}$ and stop all enrollment. If $Z_{1,k}>u_{1,k}$, reject $H_{01}$ as well. 
% \item (2) Assess Futility in the Combined Population: Else, if $Z_{C,k} ≤ l_{C,k}$, stop enrolling from subpopulation $2$ in all future stages. In this case when $Z_{1,k}>l_{1,k}$, the following additional steps must be done:
% %(stop enrolling in all future stages including current one? What does that mean?)
% 	\subitem  (a) If $Z_{1,k} > u_{1,k}$, we reject $H_{01}$ and stop all enrollment.
% 	\subitem  (b) If $Z_{1,k} ≤ l_{1,k}$, we fail to reject either $H_{0C}$ or $H_{01}$, and stop all enrollment.
% 	\subitem  (c) Else, we continue to enroll from subpopulation $1$, and re-evaluate steps (2)-(3) at the end of the next stage. 
% \item  (3) If $l_{C,k} < Z_{C,k} ≤ u_{C,k}$, continue enrolling from both subpopulations.
% \end{description}

\section{Related software}
\label{ADDPLAN}

The most comparable available software tools are AptivSolutions ADDPLAN PE (Participant Enrichment), and the \pkg{asd} \proglang{R} package \citep{parsons2012r}. Both have features that our software does not have. Conversely, there are features of our software that ADDPLAN PE and \pkg{asd} do not have.

ADDPLAN PE is a versatile, commercial software tool that implements many types of adaptive enrichment designs. One limitation is that the user must a priori designate a particular stage (e.g., stage 2) at which a change to enrollment may be made, even though there may be large prior uncertainty as to when sufficient information will have accrued to make such a decision. In contrast, \pkg{interAdapt} is more flexible, in that one can select designs in which the decision to change enrollment criteria may occur at any stage (by setting $k^*$ to the maximum number of stages $K$). 
 \pkg{interAdapt}  implements the class of designs from \citep{Rosenblum2013AdaptMISTIE}, while ADDPLAN PE does not. However, ADDPLAN implements a wide variety of other decision rules and testing procedures not available in  \pkg{interAdapt}.
Finally, \pkg{interAdapt}  is cross-platform and open-source, while ADDPLAN PE is commercial software that is only compatible with the Windows OS.

The \pkg{asd} package allows \proglang{R} users to generate  two-stage adaptive designs, which can be used to combine phase II and phase III clinical trials into a seamless design \citep{parsons2012r}. Unlike \pkg{interAdapt}, the \pkg{asd} package can generate  not only adaptive enrichment designs, but also adaptive designs that test multiple different treatments. Also in \pkg{asd}, decision rules at interim analyses can be based on short-term outcomes for each subject enrolled, if the long-term outcome is not yet available. However, \pkg{asd} does not allow more than two stages, unlike  \pkg{interAdapt} which allows up to 20 stages (though in practice fewer stages will probably be used, e.g., 5 stages). \pkg{asd} is an \proglang{R}  package, so is cross-platform and open-source. Using \pkg{asd}
 requires a working knowledge of \proglang{R}, while the GUI for \pkg{interAdapt} can be run in a web browser, with little to no interaction with the \proglang{R} console (see Section \ref{sec:running-interAdapt}), and so does not require knowledge of the \proglang{R} language.

\section{Running interAdapt }
\label{sec:running-interAdapt}

The \pkg{interAdapt} \proglang{R} package contains an interactive web browser application, as well as a command line function which performs the same computations. The browser application is built on the \pkg{shiny} package \citep{shiny2013manual}, with the back-end calculations done in \proglang{R}.

To access the \pkg{shiny} application, \pkg{interAdapt} requires the user's default web browser to be set to either Firefox (\url{http://www.mozilla.org}) or Chrome (\url{http://www.google.com/chrome/}). Users can then run the application either by installing \proglang{R} and the \pkg{interAdapt} package locally on their computer, or by simply using Firefox or Google Chrome to view \pkg{interAdapt} online. Both options are free and quick to set up. However, because the online application will slow down noticeably if accessed by multiple users simultaneously, we encourage heavy users to install \pkg{interAdapt} locally.

\subsection{Running interAdapt over the web}
\label{sub:running-online}

\pkg{interAdapt} is currently hosted on the RStudio webserver, and can be accessed at:\\
\url{http://spark.rstudio.com/mrosenblum/interAdapt}
%!!! General note to AF:  need to keep this link updated!

\subsection{Running interAdapt locally}
\label{sub:running-locally}

To run \pkg{interAdapt} locally, one must first install the \proglang{R} programming language. \proglang{R} runs on both Windows \& MacOS, is available for download at (\url{http://www.r-project.org/}). After downloading and installing \proglang{R}, activating the \proglang{R} application will open an ``\proglang{R} console'' window where typed commands are executed by \proglang{R}. The \pkg{interAdapt} \proglang{R} package can be installed by typing the lines below into the \proglang{R} console, while connected to the internet. The return key must be pressed after each line of code. The first and third lines will cause \proglang{R} to give feedback on the installation's progress, which we do not show here.

\vspace{5 mm}
\begin{verbatim}
install.packages('devtools')
library('devtools')
install_github(username='aaronjfisher',repo='interAdapt',subdir='r_package')
\end{verbatim}
\vspace{5 mm}

Once \pkg{interAdapt} has been installed, the \pkg{shiny} application can be run without an internet connection by opening the \proglang{R} console and typing

\vspace{5 mm}
\begin{verbatim}
library('interAdapt')
runInterAdapt()
\end{verbatim}
\vspace{5 mm}

The same calculations can be done directly in the \proglang{R} console, using the \verb!compute_design_performance! function. Further details are provided in the \pkg{interAdapt} package documentation, which can be accessed by typing \verb!help(package=interAdapt)! into the \proglang{R} console. 
The function's arguments are the same as the parameters available in the
\pkg{shiny} application (see Section \ref{sub:inputs}).
The function's value contains the output tables of the \pkg{shiny} application.
These tables can be used to generate the plots made by the application
(see Section \ref{sub:outputs}).

\section{User interface for the shiny application}
\label{sec:UI}

In this section, and in Section \ref{sec:example}, we will generally use the term \pkg{interAdapt} to refer to the \pkg{interAdapt} \pkg{shiny} application, although the inputs and outputs of the command line interface (see Section \ref{sub:running-locally}) have the same interpretation.

\begin{figure}[h]
\centering{}\includegraphics[width=1\textwidth]{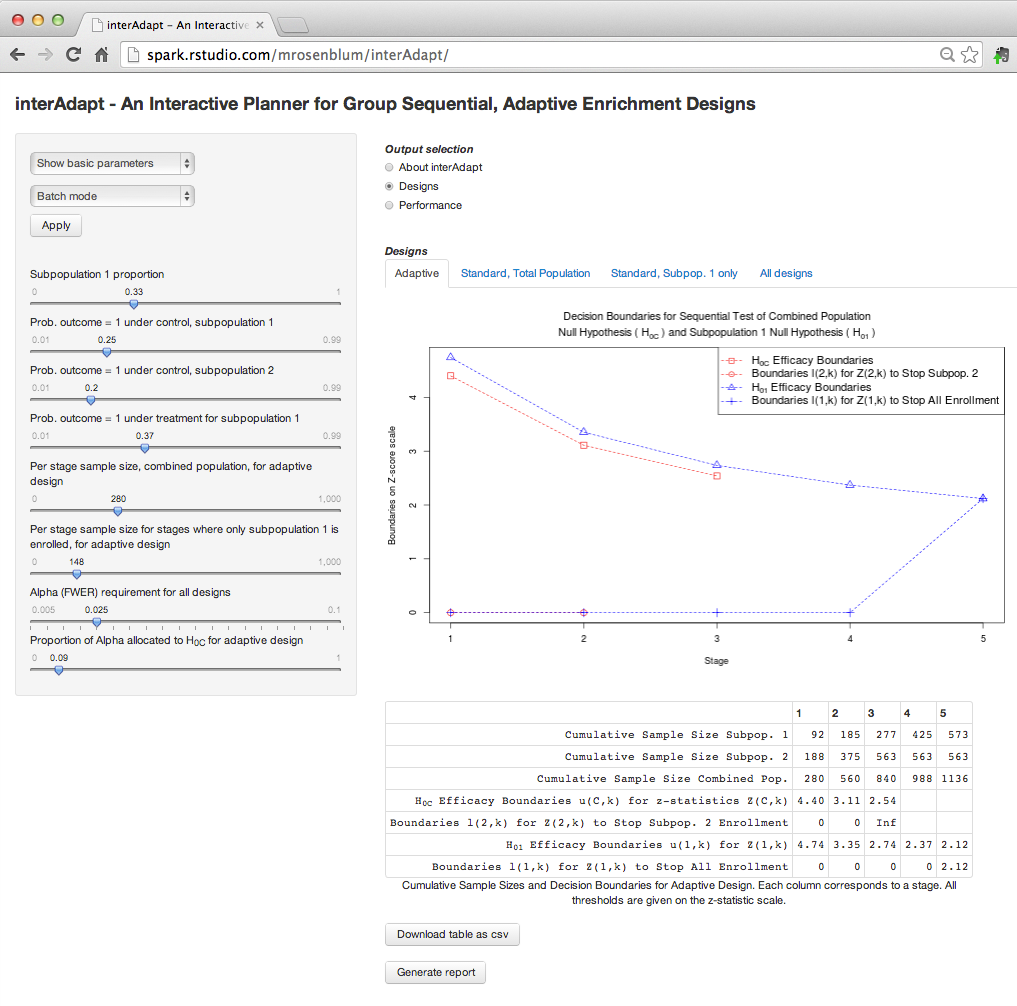} 
\caption{Designs Screenshot: Inputs can be entered in the side panel on the left, with results visible in the main panel on the right. The drop down menus at the top of the side panel can be used to navigate different interfaces to input parameters. Here we show the ``Basic parameter'' inputs, in ``Batch mode,'' where the Apply button must be pressed to update the results in the main panel. The radio buttons at the top of the main panel can be used to navigate between design outputs describing the decision rules for each trial, and performance summaries for each trial. In this figure we show the design for the adaptive trial ($AD$), based on the default input parameters. Boundaries for the z-statistics $Z_{1,k}$, $Z_{2,k}$ and $Z_{C,k}$ are shown both in the plot, and in the table. The table also contains information on how many participants are enrolled in each stage. The scroll bar on the right of the web browser has been cropped out of this figure for the sake of increased screenshot resolution.  \label{fig:designs} }
\end{figure}

\begin{figure}[h]
\centering{}\includegraphics[width=1\textwidth]{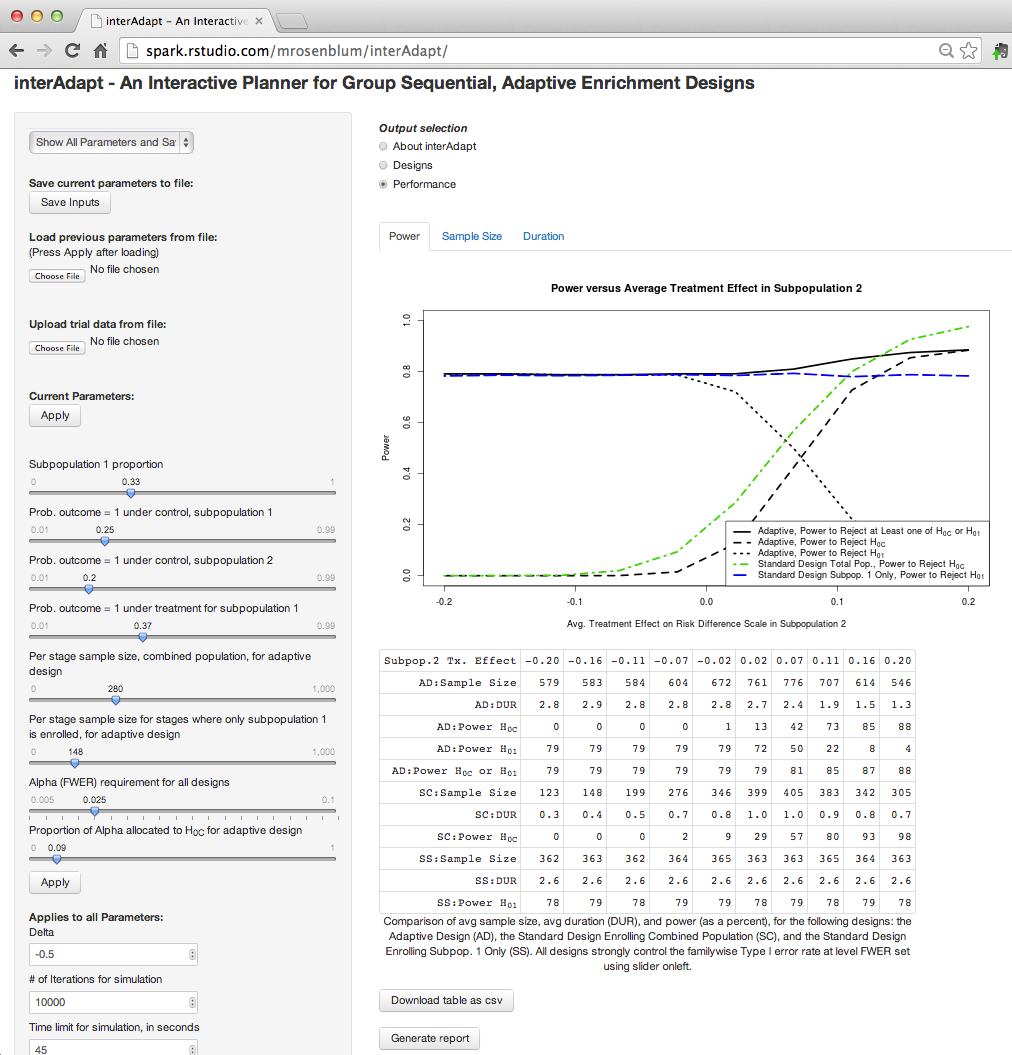} 
\caption{Performance Screenshot: Here the main panel shows performance output based on the default parameter inputs. The tabs at the top of the Performance section can be used to navigate between displays of power, expected sample size, and expected trial duration for all three designs. In the side panel, we show the interface for saving and loading sets of parameters (section \ref{sub:inputs}). Users can save the current set of inputs, load a previously used set of inputs, or upload a datafile containing results from a previous trial. If results from a previous trial are uploaded, \pkg{interAdapt} will automatically compute relevant input parameters based on this file. Additional input parameters in the side panel are available by scrolling down. As in Figure \ref{fig:designs}, the scroll bar on the right of the web browser has been cropped out of this figure for the sake of increased screenshot resolution.  \label{fig:performance} }
\end{figure}

Inputs to \pkg{interAdapt} can be entered in the side panel on the left, with outputs shown in the main panel on the right (Figures \ref{fig:designs} and \ref{fig:performance}). The parameters in the input panel let the user describe characteristics of their study populations, such as the proportion of participants in each subpopulation. 
The user can also input design requirements such as  the  familywise Type I error rate. Also, the user can input conjectured rates of success under treatment and control, to determine how well different designs perform at a given set of such values. Specifically, the user can input values for $p_{1t},p_{1c}$, and $p_{2c}$, and \pkg{interAdapt}  will compare the performance of different designs  over a range of values of $p_{2t}$, as further described below.

The main panel displays the decision boundaries and trial designs  computed by \pkg{interAdapt} to satisfy the requirements specified by the user  (Figure \ref{fig:designs}). It also compares the performance of the three designs, $AD$, $SC$ and $SS$ (Figure \ref{fig:performance}). Performance is compared in terms of power, expected sample size, and expected trial duration.

All tables generated by \pkg{interAdapt} can be downloaded as csv files by clicking on the ``Download'' button beneath the table. Users can also download a printable, html-based report of the results by clicking the ``Generate Report'' button at the bottom of the main panel (Figures \ref{fig:designs} and \ref{fig:performance}). This report is generated with the \pkg{knitr} package for \proglang{R} \citep{knitr}. Citations in the report are created using the \pkg{knitcitations} package \citep{knitcitations}.

\subsection{Inputs}
\label{sub:inputs}

Parameters in the input panel are organized into the following two sections: Basic Parameters and Advanced Parameters. To view the different sets of parameters, click the drop-down menu titled ``Show Basic Parameters.'' 

Basic Parameters can be entered using either ``Batch mode'' or ``Interactive mode''. In Batch mode, \pkg{interAdapt} will not analyze the entered parameters until the ``Apply'' button is pressed. This allows several parameters to be changed at once without waiting for \pkg{interAdapt} to recalculate the results after each individual change. In Interactive mode, \pkg{interAdapt} will automatically recalculate the results after each change, allowing the user to quickly see the effect of changing a single input parameter. Switching between Batch mode and Interactive mode can be done using the dropdown menu at the top of the Basic Parameters section. Interactive mode is not available when entering Advanced Parameters.

To save the current set of inputs, click the dropdown menu titled ``Show Basic Parameters'' and select ``Show All Parameters and Save/Load Option". You can then save the current parameters as a csv file, or load a previously saved csv file of inputs (Figure \ref{fig:performance}). Regardless of whether \pkg{interAdapt} is being run online or locally, these saved csv files are always stored on the user's computer. 

You may load a 3-column dataset into \pkg{interAdapt} in csv format, e.g., from a previous trial or study, to use in setting population parameters for simulations of hypothetical, future trials. E.g., if one is planning a Phase III trial, one might upload Phase II trial data that is already available. 
The purpose of this feature is to allow the data generating mechanisms in the \pkg{interAdapt} simulations to mimic properties of real datasets relevant to the study being planned.
 \pkg{interAdapt} 
will compute the empirical values of 
$π_1$, $p_{1c}$, $p_{1t}$, $p_{2c}$, and $p_{2t}$ from the given dataset, and set the corresponding slider values to match these.
The dataset must be structured to have one row for each participant. The first column must contain binary indicators of subpopulation, where 1 denotes subpopulation 1, and 2 denotes subpopulation 2. The second column must contain an indicator of the treatment arm ($T_i$), and the third column must contain the binary outcome measurement ($Y_i$). The first row of this dataset file is expected to be a header row of labels, rather than values for the first individual. From this dataset, \pkg{interAdapt} will calculate $π_1$, $p_{1c}$, $p_{1t}$, $p_{2c}$, and $p_{2t}$, and adjust the input sliders accordingly.
The user can then modify these parameter settings to determine how robust a given design is to differences between what was observed in previous studies and a hypothetical, future study.

A detailed explanation of each input is given below.

\subsubsection{Basic Parameters (with corresponding variables in parentheses, where applicable)} 
\label{sub:basic-params}
\begin{itemize} 

\item Subpopulation $1$ proportion ($π_1$): The proportion of the population in subpopulation $1$. This is the subpopulation in which we have prior evidence of a stronger treatment effect. 

\item Probability outcome = 1 under control, subpopulation $1$ ($p_{1c}$): The probability of a successful outcome for subpopulation $1$ under assignment to the control arm. This is used in estimating power and expected sample size of each design.

\item Probability outcome = 1 under control, subpopulation $2$ ($p_{2c}$): The probability of a successful outcome  for subpopulation $2$ under assignment to the control arm. This is used in estimating power and expected sample size of each design.

\item Probability outcome = 1 under treatment for subpopulation $1$ ($p_{1t}$): The probability of a successful outcome for  subpopulation $1$ under assignment to the treatment arm. Note that the user does not specify $p_{2t}$; instead, \pkg{interAdapt} considers a range of possible values of $p_{2t}$ that can be set through the Advanced Parameters described below.

\item Per stage sample size, combined population, for adaptive design ($n^{(1)}$): Number of participants enrolled per stage in $AD$, whenever both subpopulations are being enrolled.

\item Per stage sample size for stages where only subpopulation 1 is enrolled, for adaptive design ($n^{(2)}$): The number of participants required for each stage in AD after stage $k^*$ (only used if $k^* < K$). For stages up to and including stage $k^*$, the number of participants enrolled from subpopulation 1 is equal to $\pi_1 n^{(1)}$.

\item Alpha (FWER) requirement for all designs ($α$): The familywise Type I error rate defined in Section~\ref{sub:typeIerror}. 
%In $AD$, this is the probability of falsely rejecting either $H_{0C}$ or $H_{01}$. In $SC$ it is the probability of falsely rejecting $H_{0C}$. In $SS$ it is the probability of falsely rejecting $H_{01}$.

\item Proportion of Alpha allocated to H0C for adaptive design ($a_C$): This is used in the algorithm in Section~\ref{sub:decisionRules} to construct efficacy boundaries for the design AD.
%To control the familywise Type I error rate in the $AD$ design, the test of $H_{0C}$ is first calibrated to have a Type I error rate equal to $a_Cα$. The decision rules for $H_{01}$ are then calibrated so that the overall familywise Type I error rate is equal to $α$.

\end{itemize}

\subsubsection{Advanced Parameters (with corresponding variables in parentheses, where applicable)}
\label{sub:advanced-parameters}

\begin{itemize}

\item Delta (δ): This parameter is used as the exponent in defining the efficacy and futility boundaries as described in Section~\ref{sub:decisionRules}. %which are all proportional to $\{(\sum_{k'=1}^{K} n_{k'})/(\sum_{k'=1}^{k}n_{k'})\}^{-δ}$. 

\item \# of Iterations for simulation: This is the number of simulated trials used to 
 approximate the power, expected sample size, and expected trial duration. In each simulated trial,
 z-statistics are simulated from a multivariate normal distribution (determined by the input parameters).
The greater the number of iterations, the more accurate the simulation results will be.
It is our experience that a simulation with 10,000 iterations takes about 7-15 seconds on a commercial laptop.

\item Time limit for simulation, in seconds: If the simulation time exceeds this threshold, calculations will stop and the user will get an error message saying that the application has ``reached CPU time limit''. To avoid this, either the number of iterations can be reduced, or the time limit for the simulation can be extended. \pkg{interAdapt} does not allow for the time limit to exceed 90 seconds in the online version; there is no such restriction on the local version.

\item Total number of stages ($K$): The total number of stages, which is used in each type of design. The maximum allowed number of stages is 20.

\item Last stage subpopulation $2$ is enrolled under adaptive design ($k^*$): In the adaptive design, no participants from subpopulation $2$ are enrolled after stage $k^*$. 

\item Participants enrolled per year from combined population: This is the assumed enrollment rate (per year) for the combined population. It impacts the expected duration of the different trial designs. The enrollment rates for  subpopulations $1$ and $2$ are assumed to equal the combined population enrollment rate multiplied by $π_1$ and $π_2$, respectively. I.e., enrollment rates are proportional to the relative sizes of the subpopulations. This reflects the reality that enrollment will likely be slower for smaller subpopulations.
Active enrollment from one subpopulation is assumed to have no effect on the enrollment rate in the other subpopulation. This implies that each stage of the $AD$ design up to and including stage $k^*$ takes the same amount of time to complete, regardless of whether enrollment stops for subpopulation 2. Also, each stage after $k^*$ takes the same amount of time to complete.

\item Per stage sample size for standard group sequential design ($SC$) enrolling combined pop. ($n_{SC}$): The number of participants enrolled in each stage for $SC$.

\item Per stage sample size for standard group sequential design ($SS$) enrolling only subpop. 1 ($n_{SS}$): The number of participants enrolled in each stage for $SS$.

\item Stopping boundary proportionality constant for subpopulation 2 enrollment for adaptive design ($f_{AD,2}$): This is used to calculate the futility boundaries ($l_{2,k})$ for the z-statistics calculated in subpopulation 2 ($Z_{2,k}$) as defined in Section~\ref{sub:decisionRules}.

\item $H_{01}$ futility boundary proportionality constant for the adaptive design ($f_{AD,1}$):  This is used to calculate the futility boundaries ($l_{1,k}$) for the z-statistics calculated in subpopulation 1 ($Z_{1,k}$) as defined in Section~\ref{sub:decisionRules}.

\item $H_{0C}$ futility boundary proportionality constant for the standard design ($f_{SC}$): This is used to calculate the futility boundaries for $H_{0C}$ in $SC$ as defined in Section~\ref{sub:decisionRules}. 

\item $H_{01}$ futility boundary proportionality constant for the standard design ($f_{SS}$):  This is used to calculate the futility boundaries for $H_{01}$ in $SS$ as defined in Section~\ref{sub:decisionRules}. 

\item Lowest value to plot for treatment effect in subpopulation 2: \pkg{interAdapt} does simulations under a range of treatment effect sizes $p_{2t}-p_{2c}$ for subpopulation $2$. This sets the lower bound for this range. This effectively sets the lower bound for $p_{2t}$, since $p_{2c}$ is set by the user as a Basic parameter.

\item Greatest value to plot for treatment effect in subpopulation 2: \pkg{interAdapt} does simulations under a range of treatment effect sizes $p_{2t}-p_{2c}$ for subpopulation $2$. This sets the upper bound for this range.

\end{itemize}

\subsection{Outputs}
\label{sub:outputs}

The output panel on the right side of the user interface is split into the following three sections: ``About interAdapt'', ``Designs'', and ``Performance.'' Users can navigate between these sections using the radio buttons at the top of the panel. The About interAdapt section gives a brief introduction to the software, and a link to the full software documentation. The Designs section describes the design parameters for each of the three trials: $SC$, $SS$, and $AD$. This includes the efficacy and futility boundaries, and the maximum cumulative number of participants enrolled by the end of each stage (under no early stopping). The Performance section compares the three designs in terms of their power, expected sample size, and expected duration. 

\subsubsection{Designs}
\label{sub:design}

The Designs section gives design features that result from the user's inputs. Tabs at the top of the page can be used to navigate between the different designs. Each of the first three tabs  corresponds to one of the designs, and the fourth tab shows all three together.

In the ``Adaptive'' tab, the table at the bottom of the page shows the maximum cumulative number of participants enrolled by the end of each stage for the design AD. This is broken down by subpopulation. For each stage $k$, the table also gives efficacy boundaries for $Z_{1,k}$ and $Z_{C,k}$, and futility boundaries for $Z_{1,k}$ and $Z_{2,k}$. Because AD  stops enrolling subpopulation $2$ after stage $k^*$, futility boundaries $l_{2,k}$ (for statistics $Z_{2,k}$) in stages $k>k^*$  are not given, and $l_{2,k^*}$ is set to ``Inf" (indicating $\infty$). % which means guaranteed stopping of subpopulation 2 for futility at stage $k^*$. 
Efficacy boundaries for $Z_{C,k}$ are not given for stages $k>k^*$ since by construction (see Section~\ref{sub:decisionRules}) the AD design does not test $H_{0C}$ after stage $k^*$. (It is an area of future research to consider designs that continue to test $H_{0C}$ even after enrollment for subpopulation 2 has stopped.) A plot at the top of the page displays the efficacy and futility boundaries over all stages of the trial.

The two tabs for the standard designs SC and SS have a comparable layout to that for AD. Note that the efficacy boundaries for $SS$ and $SC$ are identical. This is because the efficacy boundaries for SC and SS are both proportional to $(k/K)^\delta$, with proportionality constants set to achieve Type I error $\alpha$, which leads to identical proportionality constants for SC and SS.

The final tab combines the tables from the first three tabs, and omits plots of the decision boundaries.

\subsubsection{Performance output}
\label{sub:performance-output}

\pkg{interAdapt} displays the performance of each of the three designs in terms of three metrics: power, expected sample size, and expected duration. These metrics all depend, among other things, on the true treatment effect in each subpopulation. A treatment effect for subpopulation $1$ can be specified in the Basic Parameters section, and a range of values for the treatment effect in subpopulation $2$ can be specified in the Advanced Parameters section. \pkg{interAdapt} calculates performance metrics (using simulations as described in the section on Advanced Parameters) for the specified range of treatment effects, and generates plots  of each metric versus the treatment effect in subpopulation $2$. The plot showing each metric can be accessed via the tabs at the top of the page. The table at the bottom of the Performance section shows all three metrics, with each column of the table denoting a different treatment effect in subpopulation $2$.

The power plot shows the power of $AD$ to reject $H_{0C}$, to reject $H_{01}$, and to reject at least one of $H_{0C}$ or $H_{01}$. Since the standard design $SC$ only tests $H_{0C}$, \pkg{interAdapt} only shows its power to reject $H_{0C}$. Similarly, \pkg{interAdapt} only shows the power of $SS$ to reject $H_{01}$. 
The power of $SS$ is constant with respect to the true treatment effect in subpopulation $2$. This is as expected, since $SS$ does not enroll any participants from subpopulation $2$. 

For the standard designs $SC$ and $SS$, the expected duration is proportional to the expected sample size. 
However, for the AD design, this does not hold; this is because the total trial duration is not necessarily proportional to the total sample size. E.g., stopping subpopulation 2 will reduce the sample size but not necessarily reduce the trial duration if subpopulation 1 is not stopped.

\section{Example of entering input and interpreting output}
\label{sec:example}

The default inputs to \pkg{interAdapt} come from the motivating example of planning the MISTIE Phase III trial. We next summarize the design goals of the investigators, based on \citep{Rosenblum2013AdaptMISTIE}. The MISTIE III trial aims to assess a new surgical treatment for stroke. The primary outcome is based on
 each participant's  disability score on the modified Rankin Scale (mRS). A successful outcome was defined as a mRS score less than or equal to 3. 

At the time of planning the Phase III MISTIE trial, the previous Phase II trial had only enrolled participants with little or no intraventricular hemorrhage (IVH). More specifically, participants had been categorized as ``small IVH'' if their IVH volume was less than 10ml, and did not require a catheter for intracranial pressure monitoring. Otherwise, participants were classified as ``large IVH.'' The Phase II trial only recruited small IVH participants, and yielded a treatment effect estimate of 12.1\% [95\% CI: (-2.7\%, 26.9\%)]. The investigators thought that the treatment could also be effective in large IVH participants, but very little data was available to assess this. We refer to those with small IVH as subpopulation $1$, since there was more prior evidence of treatment efficacy in this subpopulation; those with large IVH  are subpopulation 2.

The study designers focused on the following three scenarios of special interest:

\begin{description}
\item  (a) The average treatment effect (on the risk difference scale) is $12.5\%$ for both small and large IVH participants;
\item  (b) The average treatment effect is $12.5\%$ for small IVH participants, and zero for large IVH participants;
\item  (c) The treatment effect is zero for both subpopulations. 
\end{description}

The goals were as follows:

\begin{description}
\item  (i) At least 80\% power for testing $H_{0C}$ in scenario (a);
\item  (ii) At least 80\% power for testing $H_{01}$ in scenario (b);
\item  (iii) Familywise Type I error rate (α) of $0.025$.
\end{description}
Furthermore, the familywise Type I error rate was to be strongly controlled at level $0.025$.

Based on prior research by \cite{Hanley2012}, the proportion of participants with small IVH ($π_1$) was projected to be 0.33, the probability of a positive outcome under control was projected to be 0.25 for small IVH participants ($p_{1c}$), and the probability of a positive outcome under control was projected to be  0.2 for large IVH participants ($p_{2c}$).  If the true average treatment effect in subpopulation $1$ is 12.5\%, then the probability of a positive outcome under treatment for participants in subpopulation $1$ ($p_{1t}$) is projected to be 12.5\%+25\%=37.5\%.

Parameters for the adaptive design $AD$ were computed to achieve all three goals (i)-(iii), as fully described by \citep{Rosenblum2013AdaptMISTIE}. The corresponding standard designs, used for comparison, only had to satisfy subsets of these goals. This is to show the cost of achieving all three goals (since the adaptive design generally requires greater expected sample size, in return for achieving all three goals instead of a subset of the goals). 
The standard design $SC$ was set to achieve goals (i) and (iii), and the standard design $SS$ was set to achieve (ii) and (iii). Recall that \pkg{interAdapt} allows the user to specify a range of treatment values for subpopulation $2$, and will display the power of the trial designs across this range. By default, \pkg{interAdapt} sets the range of values for the mean treatment effect in subpopulation $2$ to be [-0.2, 0.2]. This includes scenarios (a) and (b) since in scenario (a) the mean treatment effect in subpopulation 2 is 0.125, and in scenario (b) the mean treatment effect is $0$. 

The remaining default input parameters for the $AD$ design are based on the adaptive enrichment design in Section~5.2 of \citep{Rosenblum2013AdaptMISTIE}. 
They constructed this design by first setting  $K=5$ and $δ=-.5$, and then searching over a large class of parameter values with the goal of minimizing the average expected sample size over scenarios (a)-(c), while still achieving goals (i)-(iii). They found a minimum average expected sample size at $k^*=3$, $n^{(1)}=280$, $n^{(2)}=148$, $a_C = .09$, and $f_{AD,2}=f_{AD,1}=0$.

Now we turn to the output of \pkg{interAdapt} that results from the default parameters, and show that each of the three designs achieves its corresponding goals. In the power plot, we see that $AD$ has 80\% power to reject $H_{0C}$ in scenario (a), and 80\% power to reject $H_{01}$ in scenario (b). $SC$ has 80\% power to reject $H_{0C}$ in scenario (a), and $SS$ has 80\% power to reject $H_{01}$ in scenario (b) (Figure \ref{fig:performance}). Although it is not shown,  the familywise Type I error rate is at most  .025, as this was the specified value of $\alpha$ input to \pkg{interAdapt}, and the designs are guaranteed to strongly control the familywise Type I error rate at level $\alpha$ by
 Theorem~5.1 of  \citep{Rosenblum2013AdaptMISTIE}.

\section*{Summary}
\label{sec:Summary}

We described the \pkg{interAdapt} \proglang{R} package and \pkg{shiny} application for designing and simulating trials with adaptive enrollment criteria. We provided an overview of the theoretical problem the application addresses, and gave an explanation of the application's inputs and outputs.

Current limitations of the software include that the outcome is assumed to be binary. We also currently only consider the case where outcomes are measured without delay, immediately after participants are enrolled. Relaxing both of these requirements is a goal of future work.

\section*{Acknowledgements}
\label{sec:acknowledgements}
This research was supported by U.S. National Institute of Neurological Disorders and Stroke (grant numbers 5R01 NS046309-07 and 5U01 NS062851-04), the U.S. Food and Drug Administration through the ``Partnership in Applied Comparative Effectiveness Science,'' (contract HHSF2232010000072C), and the National Institute of Environmental Health Sciences (grant number T32ES012871). This publication's contents are solely the responsibility of the authors and do not necessarily represent the official views of the above agencies.

\bibliographystyle{plainnat}
\bibliography{interAdapt}

\end{document}